\documentclass[twocolumn,showpacs,prl]{revtex4}

\usepackage{graphicx}%
\usepackage{dcolumn}
\usepackage{amsmath}
\usepackage{latexsym}

\begin {document}

\title
{
 Percolation in Models of Thin Film Depositions
}
\author
{
N. I.  Lebovka$^1$, S. S. Manna$^{2,3}$, S. Tarafdar$^4$ and N. Teslenko$^1$
}
\affiliation
{
$^1$F. D. Ovcharenko Biocolloid Chemistry Institute, 42 Vernadsky Av., Kyiv, Ukraine\\
$^2$Satyendra Nath Bose National Centre for Basic Sciences
Block-JD, Sector-III, Salt Lake, Kolkata-700098, India\\
$^3$The Abdus Salam International Centre for Theoretical Physics, Strada Costiera 11,
    I-34100 Trieste, Italy \\
$^4$Condensed Matter Physics Research Centre, Physics Department,
Jadavpur University, Kolkata 700032, India}
\begin{abstract}
We have  studied the percolation behaviour of deposits for
different (2+1)-dimensional models of surface layer formation. The
mixed model of deposition was used, where particles were deposited
selectively according to the random (RD) and ballistic (BD)
deposition rules. In the mixed one-component models with
deposition of only conducting particles, the mean height of the
percolation layer (measured in monolayers) grows continuously from
0.89832 for the pure RD model to 2.605 for the pure RD model, but
the percolation transition belong to the same universality class,
as in the 2- dimensional random percolation problem. In two-
component models with deposition of conducting and isolating
particles, the percolation layer height approaches infinity as
concentration of the isolating particles becomes higher than some
critical value. The crossover from 2d to 3d percolation was
observed with increase of the percolation layer height.
%The density of the conducting particles
%in a percolation layer shows a clear scaling behaviour versus
%percolation layer height both for BD and RD models.
\end{abstract}
\rightline{}
\maketitle

%61.43.Hv|Fractals; macroscopic aggregates (including diffusion-limited aggregates)
%64.60.Ak|Renormalization-group, fractal, and percolation studies of phase transitions|
%68.00.00|Surfaces and interfaces; thin films and low-dimensional systems

%72.60.+g|Mixed conductivity and conductivity transitions
%81.15.Aa|Theory and models of film growth
%89.75.Da|Systems obeying scaling laws, 89.00.00|Other areas of applied and interdisciplinary physics

The thin film formation processes by deposition of particles
on a substrate are of great interest both from theoretical as well as
experimental point of view \cite {Family91,Barabasi95}. The
different aspects of this problem are important in the technical
applications for production of thin-film devices, metal-insulator
mixture films, composite films with specific physical properties,
etc. \cite{Technical}.

The rather important field of investigation is related to the
electrical conductivity of thin films, which depends strongly on
their morphology and microstructure. Many works were devoted to
investigations of the fractal, percolation and electrical
properties of thin films and deposits \cite{Percolation}. It was
shown that the percolation transition in very thin (quasi
2-dimensional) films belongs to the same universality class as in
the random percolation problem \cite {Voss81, Silverman94}. The
film electrical conductivity shows also a clear transition from
the two-dimensional to three-dimensional behaviour with film
thickness increase \cite{2d3d, Kapitulnik83}. Some correlations
were observed between the conductivity and porosity for deposits
grown in a model of ballistic deposition \cite{Tarafdar00}. Jensen
et al. \cite{Jensen95} and Family \cite{Family99} investigated in
their numerical simulation works the percolation behavior for
different models of submononolayer deposits on two-dimensional
substrates.

The purpose of this work is to study the percolation behaviour for
different lattice models of three dimensional deposits growing on
the plane substrates. The spanning cluster forms in the substrate
plane. The percolation in deposits has a correlated character,
because the sites of lattice get filled dynamically during the
growth in accordance with the deposition rules.

%---------------------------------------------------------------------------
\begin{figure}[top]
\begin{center}
\includegraphics[width=7.0cm]{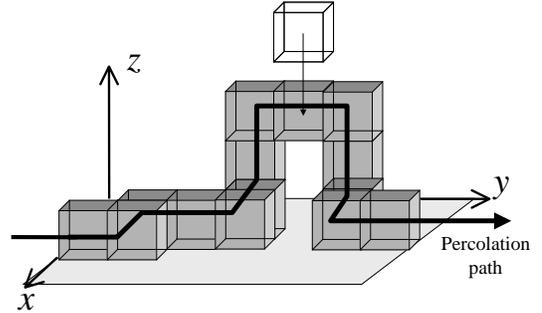}
\end{center}
\caption{ Scheme of percolation cluster formation for $2+1$
dimensional deposition model. } \label{f01}
\end{figure}
%---------------------------------------------------------------------------

In our model the particles are modeled by unit cubes. They are
deposited on an initially flat horizontal surface on the $x-y$
plane of size $L \times L$. The particles come down vertically
along the $-z$ direction with the integer $x,y$ co-ordinates and
are deposited on the substrate either by the ballistic deposition
(BD) process or by the random deposition (RD) process or by a
mixture of both the processes. In the RD process a particle comes
down vertically till it lands over a particle on the
substrate where as in the BD  process the particle gets stuck to
the substrate when any of its four vertical sides comes in contact with any
previously deposited particle of the substrate or it directly
lands on the substrate. In a mixed RD+BD model, the $s$ fraction
of particles is deposited according to the BD-rule and the
remaining $1-s$ fraction is deposited according to the RD-rules.
We call this model as: BD$_s$RD$_{1-s}$ model. The parameter $s$
allows a continuous tuning between RD model with no interaction
between particles and BD model with strong short-ranged
interaction between particles. The model of this type or other
similar models where different kind of interactions between
particles may exist are widely used for simulation of structure of
thin films with realistic morphology \cite{Mixed}

%---------------------------------------------------------------------------
%\begin{figure}[top]
%\begin{center}
%\includegraphics[width=7.0cm]{figure01_Lebovka.eps}
%\end{center}
%\caption{ Log-log plots of $\bar h_\infty - \bar h$ and $p_\infty
%- p $ versus $L$ for BD model. Here height $\bar h$ and density
%$p$ of deposit in percolation point are obtained in the limit of
%infinite dimension of lattice $L\rightarrow \infty$. Dashed lines
%correspond to the slope of $-3/4$ } \label{f01}
%\end{figure}
%---------------------------------------------------------------------------

In the mono-component model all particles are considered to be
conducting. In the bi-component model we have a $f$ fraction of
insulating particles and the $1-f$ fraction of conducting particles.

Particles are deposited on the substrate one after another and the
average height of the deposit grows. Conduction takes place
between two conducting particles when they have one surface in
contact. We stop the growth process when the deposit starts
conducting for the first time in the direction parallel to the
surface. At this percolation point, a spanning cluster across the
system is formed along the $x$ or $y$ direction (Fig.
\ref{f01}).The percolation point is easily checked by a
Hoshen-Kopelman algorithm \cite{HK}.

%The percolation point is easily checked by a modified
%Hoshen-Kopelman algorithm.
%The vertical surfaces at the boundary
%of the substrate at $x=0,L$ are numbered as 1 and 2 and that at
%$y=0,L$ are numbered as 3 and 4. A particle which is in contact
%with any of these four surfaces is assigned the same cluster
%number as the number of the surface. Therefore at an intermediate
%stage of the deposition process, we have four clusters labeled
%1,2,3 and 4 and also other clusters with labels 5 and higher which
%are not in contact with any of the four boundary surfaces. When a
%particle coalesces two arbitrary clusters of particles, the
%resultant cluster gets the smaller of the two labels. Therefore at
%the percolation point a single cluster spanning the two opposite
%vertical surfaces is formed when a particle makes a connection
%between the cluster numbers 1 and 2 or 3 and 4. At this point we
%stop the deposition process and make the measurements.

During the deposition process, the time elapsed is measured in
units of the number of equivalent complete layers deposited.
Therefore $N$ particles have been deposited in time $t=N/L^2$. On
the other hand the mean height of the deposit at time $t$ is $\bar
h=\Sigma_{x,y}h_{xy}(t)/L^2$. The percolation density is the
volume fraction $p$ of the conducting particles at the percolation
point i.e., the ratio of the number of conducting particles $N_c$
and the total volume of the deposit $p= N_c/(\bar hL^2)$. In
mono-component model $N_c=N$ and $p= t/\bar h$. For RD model the
bulk of the deposit is compact (without any pores in the vertical
columns) but it has a rough interface. Therefore the RD limit at
$s=0$ corresponds to $p=1$. For ballistic deposition (BD) model at
$s=1$ the deposit is porous and consequently $p<1$. For a
bi-component model we have a mixture of conducting and insulating
particles. Here the total density of the deposit including
conducting and insulating particles is $p_{total}=N/(\bar hL^2)$
and $N_c=N(1-f)$ and $p=t(1-f)/\bar h=p_{total}(1-f)$.

%---------------------------------------------------------------------------
\begin{figure}[top]
\begin{center}
\includegraphics[width=7.0cm]{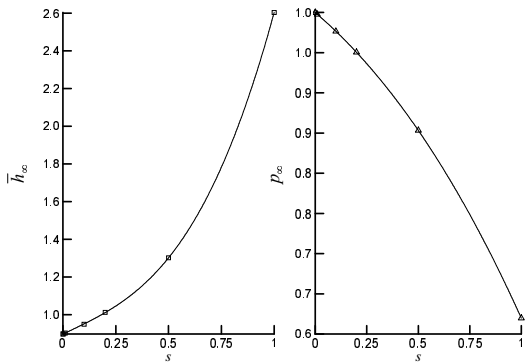}
\end{center}
\caption{ Plots of height $\bar h_\infty$ and density $p_\infty$
of deposit in percolation point versus parameter $s$ for
one-component mixed BD$_s$RD$_{1-s}$ model. The data error is of
order of data symbol size. The lines serve as a guide to the eye.
} \label{f02}
\end{figure}
%---------------------------------------------------------------------------

For finite extensions $(L)$ of the substrates the percolation
height $\bar h(L)$ and $p(L)$ are $L$ dependent. The values of
 $\bar h(L)$ and  $p(L)$
are determined for different substrate sizes $L$
%by all
%three ways. Varying $L$ we see that the percolation height is
%underestimated when $a$ condition is used and overestimated when
%$b$ condition is used as compared with the value of $\bar h(\infty)$.
%A lattice edge width $L$
varied from $8$ to $2048$ and the periodical boundary conditions
were applied in deposition rules along the directions $x$ and $y$.
Results were averaged over 100-5000 different runs, depending on
the size of the lattice and required precision.

In analogy with the corresponding finite size behaviours in the
ordinary percolation, we assume the following relations:
\begin{equation}
p(L)=p_{\infty}+a_{p}L^{-1/\nu_p}, \label{e01}
\end{equation}
and
\begin{equation}
\bar h(L)=\bar h{(\infty)}+a_hL^{-1/\nu_h}, \label{e02}
\end{equation}
where $\nu_p, \nu_h\approx 4/3$.

The probability that a particle is deposited along a particular
vertical line on the $x-y$ plane is $1/L^2$ which is very small
when $L$ is large such that the mean height $\bar h$ is maintained
at a fixed value when $N$ particles are deposited. This implies
that in the RD model the number $h$ of particles in an arbitrary
column of particles follow a Poisson distribution:
\begin{equation}
P(h)=(e^{-\bar h})({\bar h}^{h})/h!.
\end{equation}
Therefore the probability of an empty column ($h=0$) is equal to
$P(0)=e^{-\bar h}$. In percolation point
\begin{equation}
P(0)=1- p_{2d}=e^{-\bar h},
\end{equation}
where $p_{2d}$ %\approx 0.592746$ \cite {Jan99}
is the percolation
threshold for the square lattice site percolation problem.

Taking into account the finite size scaling behavior of $p$:
\begin{equation}
p_{2d}(L)=p_{2d,\infty}+a_{2d,p} L^{-1/\nu_p},
\end{equation}
where $p_{2d,\infty}=0.592746..$ is the percolation concentration
in the limit of infinite system $(L\rightarrow\infty)$ and $\nu_p
=4/3$ is a correlation length scaling exponent \cite{Stauffer92},
we obtain
\begin{equation}
\bar h(L)=-\ln(1-p_{2d,\infty}-a_{2d,p} L^{-1/\nu_p})
\end{equation}
\begin{equation}
=\bar h_\infty+\ln(1-\frac{a_{2d,p}}{(1-p_{2d,\infty})L^{1/\nu_p
}})\approx \bar h_\infty-a_{h}L^{-1/\nu_h},
\end{equation}
where $\bar h_\infty=-\ln(1-p_{2d,\infty})\approx 0.89832$,
$a_h=a_{2d,p}/(1-p_{2d,\infty})$, $\nu_h=\nu_p$.

We  see that  for pure RD model $\nu_h=\nu_p=4/3$. So, the RD
model belong to same class of universality as the 2-dimensional
random percolation model. This fact reflect the small mean height
of (2+1)-dimensional random deposit $\bar h_\infty\approx
0.89832$, which is only slightly higher than mean height of
2-dimensional random deposit $\bar h_\infty\approx 0.592746$.

%---------------------------------------------------------------------------
\begin{figure}[top]
\begin{center}
\includegraphics[width=7.0cm]{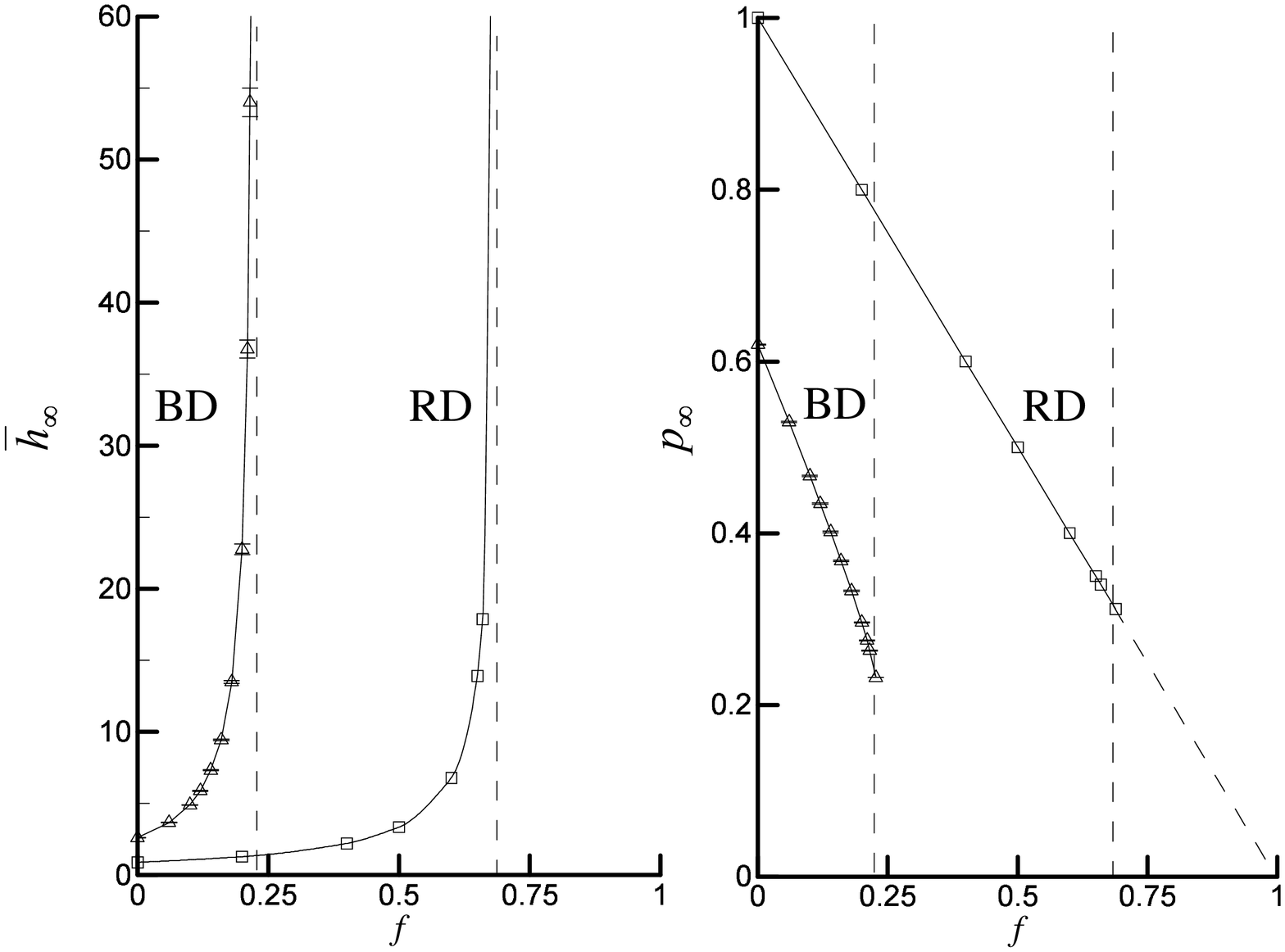}
\end{center}
\caption{ Plots of height $\bar h_\infty$ and density $p_\infty$
of deposit in percolation point versus fraction of isolating
particles for two-component BD and RD models. In cases when it is
not show directly the data error is of order of data symbol size.
The solid lines serve as a guide to the eye. Vertical dashed lines
show the critical concentrations which are  $f_c=0.227 \pm 0.001$
and $f_c=0.70 \pm 0.01$ for BD and RD models respectively. }
\label{f03}
\end{figure}
%---------------------------------------------------------------------------

%Like in the usual random percolation problem, the threshold point is
%determined in three different ways:
%{\em (a)} The deposit is conducting along either $x$ or $y$ direction,
%{\em (b)} the deposit is conducting along only $x$ direction and
%{\em (c)} the deposit is conducting along both $x$ and $y$ directions.

 On the basis of numerical simulations we estimate that for pure $BD$ at $s=1$,
$\bar h_{\infty}=2.605 \pm 0.005$ and $p_{\infty}=0.620 \pm 0.005$
in the limit $L=\infty$. Using these asymptotic values we plot
$\bar h_{\infty}-\bar h(L)$ and $p_{\infty}-p(L)$ vs, $L$ on
double logarithmic scales and %(in Fig. \ref{f01}). We see, that
in both cases plots correspond to the slope of -3/4, which means
$\nu_h=\nu_p\simeq 4/3$.

%---------------------------------------------------------------------------
\begin{figure}[top]
\begin{center}
\includegraphics[width=7.0cm]{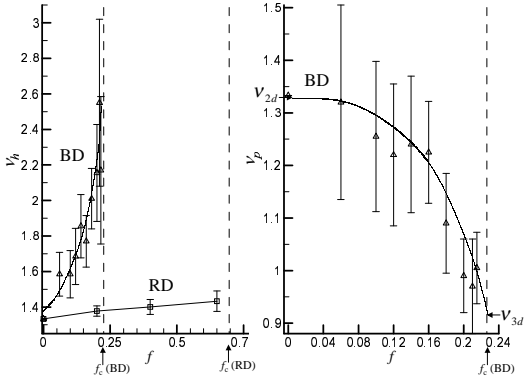}
\end{center}
\caption{ Plots of scaling exponents $\nu_h$ and $\nu_p$ in Eqs.
(\ref{e01},\ref{e02}) versus fraction of isolating particles for
two-component BD and RD models. The solid lines serve as a guide
to the eye. Vertical dashed lines show the critical concentration
which are $f_c=0.227 \pm 0.001$ for BD model and $f_c=0.70 \pm
0.01$ for BD model.} \label{f04}
\end{figure}
%---------------------------------------------------------------------------

Percolation height $\bar h_{\infty}$ and the percolation densities
$p_{\infty}$ are similarly calculated for the mixed
BD$_s$RD$_{1-s}$ model varying the mixing parameter $s$ and
plotted in Fig. \ref{f02}. The height of the deposit $\bar
h_\infty$ increases and its density decreases smoothly with
increase of the fraction of deposited BD particles. In the limit
of pure RD model, the theoretical value $\bar h_{\infty}\approx
0.89832$ is observed well. Same as both pure BD and RD models, the
mixed BD$_s$RD$_{1-s}$ also displays the scaling behaviour
described by Eqs. (\ref{e01},\ref{e02}) with scaling exponent
$\nu_h=\nu_p\approx 4/3$. Using the calculated $\bar h_\infty(s)$
and $p_\infty(s)$ dependencies and substituting $\nu_h=\nu_p =
4/3$ into Eqs. (\ref{e01},\ref{e02}) the coefficients $a_h$ and
$a_p$ versus $s$ were obtained. Both of these coefficients $a_h$
and $a_p$ increase with $s$. It is important to note that for the
pure RD model $a_p=0$. It means that there is no finite size
scaling for the RD model, as for any $L$ at $p=1$ the deposit is
compact without pores by definition.

We can conclude that the mixed BD$_s$RD$_{1-s}$ model, presumably,
belong to same class of universality as the 2-dimensional random
percolation model at any value of $s$.

Figure \ref{f03} presents the deposit height $\bar h_\infty$ and
density of conducting particles $p_\infty$ in the percolation
point estimated in the limit of $L\rightarrow \infty$ versus
fraction of isolating particles $f$. The dependencies of $\bar
h_\infty (f)$ show the typical percolation behaviour: as $f$
reaches some critical value $f_c$ the value of $\bar h_\infty$
goes to infinity; it means that there is no percolation at any
finite height of deposit. The estimated values of critical
concentrations of the isolating particles are $f_c(BD)=0.227 \pm
0.001$ for the BD model and $f_c(RD)=0.70 \pm 0.01$ for RD model.

%---------------------------------------------------------------------------
\begin{figure}[top]
\begin{center}
\includegraphics[width=7.0cm]{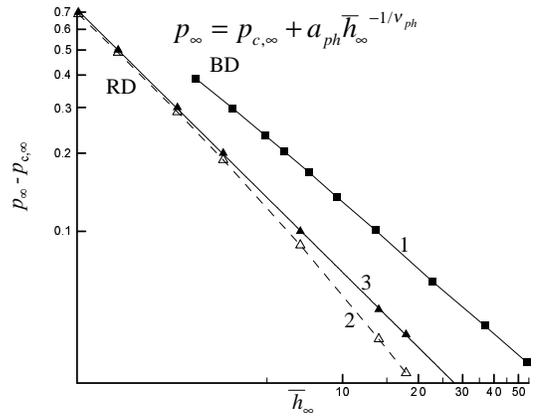}
\end{center}
\caption{Log-log plot of $p_\infty-p_{c,\infty}$ versus $\bar
h_\infty $ for two-component BD(squares) and RD(triangles) models.
See the text for the details. } \label{f05}
\end{figure}
%---------------------------------------------------------------------------

%A density of conducting particles is equal $p=p_t(1-f)$, where
%$p_t$ is the total density of conducting and isolating particles.
For the RD model, the total density of particles is $p_{total}=1$
by definition, and, so, the density of conducting particles is
$p=p_{total}(1-f)=1-f$. The linear law of $p_\infty$ decrease with
$f$ increase is actually observed in simulation data for the RD
model, this law is rather close to linear for the BD model (Fig.
\ref{f03}). In the critical point $f=f_c$, the density of
conducting particles is equal to $p_{c,\infty}=0.232\pm 0.001$ for
the BD model, and $p_{c,\infty}=0.30\pm 0.01$ for the RD model.
This value for the RD model is very close to percolation
concentration for the random percolation on a simple cubic lattice
$p=0.311609$ \cite{Jan99}. The estimated value of the total
density of deposit $p_{total,\infty}=0.300\pm 0.001$ for the BD
model coincides with the previously reported data for the deposit
density extrapolated to the infinite-system limit for the BD model
\cite{Krug91}.

The scaling exponents $\nu_h$, $\nu_p$ %and parameters $a_h$, $a_p$
obtained by the least square fit of Eqs. (\ref{e01},\ref{e02})
versus $f$ are presented at Fig. \ref{f04} %,\ref{f09}
for the BD and RD models. All fitting procedures were done at the
fixed values of $p_\infty$, two free parameters and correlation
coefficients were higher than $0.998$. For the RD model, the value
of $p$ is independent from system size $L$ by definition, so, only
$\nu_h$  dependency is presented at Fig.\ref{f04}.

The scaling exponent $\nu_h$ for the BD model continuously grows
with $s$ from $\nu_h\approx 4/3$ characteristic for single-
component model to $\nu_h\approx 2.4$ near critical point
$f=f_c=0.227$. The dependence of $\nu_h$ versus $f$  for the RD
model is less pronounced.
%The precision of $\nu_h$ determination at higher $f$ decreases, as
%the respective amplitude $a_h$ continuously grows with $s$ and
%approaches very large values at $f\rightarrow f_c$ (See Fig.
%\ref{f10}).
The correlation length exponent $\nu_p$ for BD model smoothly
decreases from $\nu_p=\nu_{2d}=4/3$ characteristic for the
2d-systems (at $f=0$) to $\nu_p=\nu_{3d}=9/10$ characteristic for
the 3d-systems (at $f=f_c=0.227$) \cite{Stauffer92}. This
behaviour can be easily understood, as far as the height of
percolation deposit at high $f$ increases and approaches the
infinity at the critical concentration of isolating particles. So,
the BD-model percolation of deposit in the vicinity of
$f=f_c=0.227$ may be considered as a percolation in a
three-dimensional system and the universality class of this
percolation model is presumably the same as for the random
percolation model. Existence of the 2d-to-3d percolation crossover
was experimentally observed in random metal-insulator mixture
films.  when thickness of the films deposited was increased
\cite{Kapitulnik83}.

As far as at $f<f_c$, ($L\rightarrow \infty)$ the height of a
percolation cluster $\bar h_\infty$ remains finite even for
infinitely large substrate dimension, it is interesting to check
$p_\infty$ versus $\bar h_\infty$ for existence of scaling:
\begin{equation}
p_\infty=p_{c,\infty}+a_{ph} \bar h_\infty^{-1/\nu_{ph}}.
\label{Scaling-ph}
\end{equation}

Figure \ref{f05} shows the log-log presentation of
$p_\infty-p_{c,\infty}$ versus versus $\bar h_\infty$ for the BD
and RD models. For the BD model, we put $p_{c,\infty}=0.232$. The
solid line $1$ corresponds to best fit of Eq.(\ref{Scaling-ph}) to
the data (filled squares) with parameters $\nu_{ph}=1.20 \pm 0.01$
and $a_{ph}=0.883 \pm 0.001$.

Putting the value $p_{c,\infty}=0.311609$ for the RD model, which
is equal exactly to the concentration for the random percolation
problem on simple cubic lattice, we obtain data represented by the
open triangles. The dashed line $2$ is drawn as a guide to the
eyes and the scaling is rather poor. But when we put a somewhat
higher value $p_{c,\infty}=0.30$ (remind that the value obtained
from $\bar h_\infty$ versus $f$ dependencies is
$p_{c,\infty}=0.30\pm 0.01$) we get the data represented by filled
triangles. Now the scaling is rather good and the solid line $3$
corresponds to best fit of Eq. (\ref{Scaling-ph}) to the data
(filled triangles) with parameters $\nu_{ph}=1.04 \pm 0.01$ and
$a_{ph}=0.633 \pm 0.004$.

In summary, we have investigated the percolation in the direction
parallel to the surface for different models of deposition layer
formation.
%We have modified existing Hoshen-Kopelman algorithm for
%labeling of growing deposit and fixing of the percolation point
%for efficient simulation of large 3d-systems.
In one-component models with deposition of only conducting
particles, the height of deposits in the point of percolation is
finite and is $\bar h_\infty=0.592746 $ for the RD model $\bar
h_\infty=2.605$ for the BD model. The mixed BD$_s$RD$_{1-s}$
model, presumably, belongs to the universality class of 2d-random
percolation problem. In two-component models with conducting and
isolating particles, the percolation layer height $\bar h_\infty$
can be varied in wide range by tuning of a concentration of the
isolating particles $f$ and a clear crossover from 2d to 3d
percolation is observed with increase of $\bar h_\infty$. The
percolation in layer is impossible when $f$ exceeds some critical
concentration $f_c$. The weakening of inter-particle interaction
results in increasing of threshold value of $f_c$, from $0.227$
for BD model  $\approx 0.7$ for RD model.

DST, Govt. of India and   Ministry of Ukraine for Education and
Science,  are gratefully acknowledged for grant of an
Indo-Ukrainian collaboration project.

\end{document}